\documentclass[11pt]{article}
\usepackage{amsmath,amssymb,amsfonts,epsf,graphicx}
\usepackage[nosort]{cite}
\usepackage[margin=1in]{geometry}

\makeatletter \@addtoreset{equation}{section} \makeatother

\newcommand{\fft}[2]{{\frac{#1}{#2}}}

\def\nn{\nonumber}

\newcommand{\be}{\begin{equation}}
\newcommand{\ee}{\end{equation}}
\def\ba{\begin{array}}
\def\ea{\end{array}}

\def\fft#1#2{\frac{#1}{#2}}

\def\sst#1{{\scriptscriptstyle #1}}

\def\dalemb#1#2{{\vbox{\hrule height .#2pt
        \hbox{\vrule width.#2pt height#1pt \kern#1pt
                \vrule width.#2pt}
        \hrule height.#2pt}}}

\newcommand{\bea}{\begin{eqnarray}}
\newcommand{\eea}{\end{eqnarray}}

\def\0{{\sst{(0)}}}
\def\1{{\sst{(1)}}}
\def\2{{\sst{(2)}}}
\def\3{{\sst{(3)}}}
\def\4{{\sst{(4)}}}
\def\5{{\sst{(5)}}}
\def\6{{\sst{(6)}}}
\def\7{{\sst{(7)}}}
\def\8{{\sst{(8)}}}

\begin{document}

\begin{center}\ \\ \vspace{60pt}
{\Large {\bf Black Five-Branes and Fluxbranes on Gravitational Instantons}}\\ 
\vspace{30pt}

Justin F. V\'azquez-Poritz
\vspace{20pt}

{\it Physics Department\\ New York City College of Technology, The City University of New York\\ 300 Jay Street, Brooklyn NY 11201, USA.}

\vspace{10pt}
{\it The Graduate School and University Center, The City University of New York\\ 365 Fifth Avenue, New York NY 10016, USA}\\

\vspace{20pt}

{\tt jvazquez-poritz@citytech.cuny.edu}

\end{center}

\vspace{30pt}

\centerline{\bf Abstract}

\noindent We apply a U-duality based solution-generating technique to construct supergravity solutions which describe nonextremal D5-branes and fluxbranes on various gravitational instantons. This includes an F7-brane wrapped on a sphere, which remains weakly-coupled in the asymptotic region. We construct various superpositions of nonextremal D5-branes and fluxbranes that have angular momentum fixed by the parameters associated with their mass and two magnetic charges.

\thispagestyle{empty}

\pagebreak



\section{Introduction and summary}

Supergravity $p$-brane solutions have played a crucial role in string theory, and a great number of them have provided examples of gauge-gravity duality \cite{agmoo}. However, if a solution contains a singularity, then this imposes a severe restriction on its range of validity. Singularities may be resolved by higher-order stringy effects or by geometrical deformations at the level of supergravity. For example, the Klebanov-Tseytlin solution describing regular and fractional 3-branes at the apex of the conifold \cite{kt} can be rendered completely regular by deforming the conifold \cite{ks}, thereby providing a supergravity description of chiral symmetry breaking and confinement. There is an abundance of work that has been done regarding the resolution of various $p$-brane solutions; see, for instance, \cite{resolution}.

Another route for dealing with a singularity is to hide it behind an event horizon. It was proposed that such a non-extremal generalization of the Klebanov-Tseytlin solution could provide a supergravity description of the restoration of chiral symmetry above a critical temperature \cite{buchel}. Due to the complexity of the equations, perturbative as well as numerical techniques have been applied for this purpose \cite{restoration1,restoration2,restoration3,restoration4}.

5-brane solutions tend to be simpler since the transverse space has only four dimensions. Resolutions of heterotic 5-branes on Eguchi-Hanson and Taub-NUT instantons have been constructed in \cite{resolution}. This resolution incorporates multiple matter Yang-Mills fields, which are only available for heterotic string theory. Nevertheless, resolutions of 5-branes in type II theories are possible if the 5-brane is wrapped around an $S^2$ \cite{wrapped5brane1} or $S^1$ \cite{wrapped5brane2} and appropriately twisted. The latter case ended up serving as a guide for the construction of an $S^1$-wrapped D3-brane on a resolved conifold \cite{s1wrappedD3}. This illustrates the usefulness of 5-brane solutions as a toy model for more complicated $p$-brane solutions with transverse spaces of higher dimensionality.

In this paper, we will consider nonextremal D5-branes on various gravitational instantons. We will apply a solution-generating technique that involves using U-duality to obtain nonextremal $p$-branes from neutral black holes \cite{sen}. Our ``seed solutions'' will be five-dimensional black holes on various gravitational instantons \cite{emparan,elvang,ford,teo1}\footnote{We do not consider black holes whose horizons are distorted lens spaces $L(n;m)=S^3/\Gamma (n;m)$ \cite{lu1}, since the resulting nonextremal 5-brane solutions would be similar to the charged generalizations already constructed in \cite{lu2}.}. We embed these solutions in eleven dimensions, perform a boost, dimensionally reduce to type IIA theory along the boosted direction, and perform a series of five T-dualities in order to obtain new nonextremal D5-brane solutions. The boost parameter in this prescription is associated with the magnetic charge of the D5-brane. Over the years, this type of prescription for generating new solutions has been used to obtain a multitude of supergravity solutions. A similar solution-generating technique has been used to generate solutions which describe the baryonic branch of the Klebanov-Strassler theory \cite{martelli}, as well as its generalization to finite temperature \cite{caceres}.

For seed solutions which have a finite and constant $S^1$ in their asymptotic region, we can incorporate a rotation, instead of a boost, in the above recipe. Reducing the eleven-dimensional solution to type IIA theory along the rotated direction then results in an F7-brane wrapped on a sphere, where the rotation parameter is associated with the magnetic charge of the fluxbrane. A fluxbrane can be thought of as a higher-dimensional generalization of the Melvin universe \cite{melvin}, which is a flux tube in four dimensions \cite{fluxtube}. Fluxbranes were introduced within the context of string theory in \cite{cosmology,costa1,saffin,gutperle,costa2}. Unlike the F7-brane found in \cite{gutperle}, the one discussed here remains weakly coupled in the asymptotic region. Applying a rotation and a boost together in the solution-generating prescription results in a superposition of a D5-brane and a smeared F2-brane (or a smeared D0-brane and an F7-brane, if one does not perform the five T-dualities), which have two independent magnetic charge parameters. These solutions have the additional feature of having angular momentum, which might seem odd given that the seed solutions do not have angular momentum. However, this angular momentum is not an independent quantity and is fixed in terms of the parameters associated with the mass and two magnetic charges.

It would be quite interesting to find an interpolation between these nonextremal 5-brane solutions and the aforementioned resolutions which either involve a heterotic 5-brane with Yang-Mills fields \cite{resolution} or a 5-brane wrapped on an $S^2$ \cite{wrapped5brane1} or $S^1$ \cite{wrapped5brane2}. An interpolating solution could provide an explicit example of the transition from a completely regular solution to one in which there is a singularity hidden behind a horizon. This could offer insight into the nature of the transition point at which a horizon develops for the more complicated case of the nonextremal generalization of the Klebanov-Strassler solution, which may not be readily apparent from perturbative or numerical techniques.

This paper is organized as follows. In section 2, we provide a couple of examples that illustrate the solution-generating technique for cases in which either a boost or a rotation is applied. The resulting solutions are a nonextremal D5-brane and an F7-brane wrapped on a sphere. In section 3, we simultaneously apply a boost and a rotation as part of the solution-generating procedure and obtain a superposition of a nonextremal D5-brane and a smeared F2-brane on a Kaluza-Klein (KK) bubble. In section 4, we present various other examples of superpositions of D5-branes and smeared F2-branes.

\section{Basic examples of the solution-generating technique}

\subsection{Nonextremal D5-brane}

The five-dimensional Schwarzschild-Tangherlini metric \cite{tangherlini} is given by 
\be
ds_5^2=-f dt^2+f^{-1} dr^2+r^2 d\Omega_3^2,
\ee
where
\be
f=1-\fft{r_0^2}{r^2}.
\ee
The event horizon is located at $r=r_0$ and the singularity is at $r=0$. The direct product of the five-dimensional Schwarzschild-Tangherlini solution with ${\mathbb R}^6$ is a vacuum solution in eleven dimensions:
\be\label{11D-metric1}
ds_{11}^2 =ds_5^2+dz^2+dx_1^2+\cdots dx_5^2,
\ee
Taking this as the seed solution, one can perform a boost in the $z$ direction,
\be\label{boost}
t\rightarrow t\cosh\beta-z\sinh\beta,\qquad z\rightarrow z\cosh\beta-t\sinh\beta,
\ee
so that the metric (\ref{11D-metric1}) can be written as
\be
ds_{11}^2 = H\left[ dz+(H^{-1}-1)\coth\beta dt\right]^2- H^{-1} f dt^2+dx_1^2+\cdots +dx_5^2+f^{-1} dr^2+r^2 d\Omega_3^2,
\ee
where
\be\label{harmonic-function}
H=1+\fft{r_0^2 \sinh^2\beta}{r^2}.
\ee
Then performing dimensional reduction along the $z$ direction yields a type IIA nonextremal D0-brane smeared along the $x_1,\dots , x_5$ directions:
\bea
ds_{10}^2 &=& -H^{-7/8} f dt^2+H^{1/8} \left( dx_1^2+\cdots +dx_5^2+f^{-1} dr^2+r^2 d\Omega_3^2\right) ,\nn\\
F_\2 &=& \coth\beta\ dH^{-1}\wedge dt,\nn\\
\phi &=& -\fft34 \log H.
\eea
If we then T-dualize along the $x_1,\dots ,x_5$ directions, we obtain the nonextremal D5-brane in type IIB theory \cite{duff}:
\bea\label{D5}
ds_{10}^2 &=& H^{-1/4}\left( -f dt^2 +dx_1^2+\cdots +dx_5^2\right)
+H^{3/4} \left( f^{-1} dr^2+r^2 d\Omega_3^2\right),\nn\\
\ast F_\3 &=& \coth\beta\ dH^{-1}\wedge dt\wedge d^5x,\nn\\
\phi &=& -\fft12 \log H.
\eea
The mass per unit 5-volume is
\be
m=\left( 2\sinh^2\beta+3\right) r_0^2,
\ee
and it has a magnetic charge 
\be
Q=r_0^2\sinh 2\beta.
\ee
Note that the event horizon of the black hole at $r=r_0$ maps into the event horizon of the D5-brane. Since the harmonic function (\ref{harmonic-function}) stays finite and nonzero for $r\ge r_0$, the source for the magnetic field strength $F_\3$ in (\ref{D5}) lies within the event horizon at $r=0$. The extremal limit of this solution can be obtained by taking $r_0\rightarrow 0$ and $\beta\rightarrow\infty$ while keeping $r_0 \sinh\beta$ constant.

\subsection{F7-brane on an $n$-sphere}

The $n+2$-dimensional Euclidean Schwarzschild-Tangherlini instanton can be embedded in eleven dimensions as
\be\label{11D-metric2}
ds_{11}^2=-dt^2+dx_1^2+\cdots dx_{7-n}^2+dz^2+f d\psi^2+f^{-1} dr^2+r^2 d\Omega_n^2,
\ee
where
\be
f=1-\left( \fft{r_0}{r}\right)^{n-1}.
\ee
The radial coordinate $r\ge r_0$ and regularity at $r=r_0$ requires that $\psi$ has the period $4\pi r_0/(n-1)$. 
This is a static $S^n$ Kaluza-Klein (KK) ``bubble of nothing" \cite{witten}. It is a direct product of Minkowski$_{8-n}$ and the Euclidean Schwarzschild instanton and is asymptotically Minkowski$_{10}\times S^1$ with a finite and constant radius for $S^1$. 

Upon performing a rotation in the $z-\psi$ plane, 
\be\label{rotation}
\psi\rightarrow \psi \cos\alpha-z\sin\alpha,\qquad z\rightarrow z \cos\alpha+\psi\sin\alpha,
\ee
the metric (\ref{11D-metric2}) can be written as
\be
ds_{11}^2 = H\left[ dz+(H^{-1}-1)\cot\alpha d\psi\right]^2- dt^2+dx_1^2+\cdots +dx_{7-n}^2+H^{-1}fd\psi^2+f^{-1} dr^2+r^2 d\Omega_n^2,
\ee
where
\be
H=1-\left( \fft{r_0}{r}\right)^{n-1} \sin^2\alpha >0.
\ee
Performing dimensional reduction along the $z$ direction yields an F7-brane wrapped on a $n$-sphere:
\bea\label{F7}
ds_{10}^2 &=& H^{-7/8} f d\psi^2+H^{1/8} \left( -dt^2+dx_1^2+\cdots +dx_{7-n}^2+f^{-1} dr^2+r^2 d\Omega_n^2\right) ,\nn\\
F_\2 &=& \cot\alpha\ dH^{-1}\wedge d\psi,\nn\\
\phi &=& -\fft34 \log H,
\eea
The total magnetic flux is
\be
\fft{1}{4\pi} \int F_{(2)}=\fft{r_0\tan \alpha}{n-1}.
\ee
Unlike the F7-brane found in \cite{gutperle}, here the IIA theory remains weakly coupled for large $r$. Note that the solution (\ref{F7}) can also be obtained by a double Wick rotation of a D0-brane smeared along $8-n$ directions. Also, as opposed to the nonextremal D5-brane, this F7-brane solution does not have a nontrivial extremal limit.

\section{Superposition of D5-brane and smeared F2-brane on KK bubble}

We will now consider a seed solution for which we can simultaneously apply a boost and a rotation as part of the solution-generating scheme. The resulting solution involves two magnetic charges parameterized by $\alpha$ and $\beta$.

The seed solution describes a black hole on the Euclidean Schwarzschild instanton, or equivalently a black hole sitting in the throat of a KK bubble, is given by the metric 
\cite{emparan}
\be\label{schwarzschild}
ds_5^2=-f dt^2+g d\psi^2+A \left( \fft{dx^2}{G(x)}-\fft{dy^2}{G(y)}+B d\phi^2\right)
\ee
where
\bea\label{Gf}
G(x) &=& (1+cx)(1-x^2),\qquad
A=\fft{2\chi^4 (1+cx)^2 (1-c)(1-y)^2}{(x-y)^3},\qquad B=-\fft{2(1+x)(1+y)}{(1-c)(x-y)},\nn\\
f &=& \fft{1+cy}{1+cx},\qquad
g = \fft{1-x}{1-y}.
\eea
The parameters $\chi$ and $c$ take the ranges $\chi>0$ and $0\le c<1$, and the $x$ and $y$ coordinates take the ranges $-1\le x\le 1$ and $-\fft{1}{c}\le y\le -1$. The horizon is located at $y=-\fft{1}{c}$ and the asymptotic region is at $x=y=-1$. For vanishing $c$, the black hole goes away and we are left with the background geometry, which is the direct product of time and the Euclidean Schwarzschild instanton and is asymptotically Minkowski$_4\times S^1$ with a finite and constant $S^1$.

We can embed this as a vacuum solution in eleven dimensions with the metric 
\be\label{general11Dmetric}
ds_{11}^2=ds_5^2+dz^2+dx_1^2+\cdots +dx_5^2,
\ee
and perform the boost (\ref{boost}) and rotation (\ref{rotation}). This yields the metric
\bea
ds_{11}^2 &=& H \left[ dz+H^{-1} (f-1) c_{\beta} s_{\beta} c_{\alpha} dt+H^{-1} (c_{\beta}^2-g-f s_{\beta}^2) c_{\alpha} s_{\alpha} d\psi\right]^2\nn\\
&-& H^{-1} K \left[ dt+K^{-1} (1-f) c_{\beta} s_{\beta} s_{\alpha} c_{\alpha}^2 d\psi\right]^2+dx_1^2+\cdots +dx_5^2\nn\\ 
&+& J d\psi^2 + A \left( \fft{dx^2}{G(x)}-\fft{dy^2}{G(y)}+B d\phi^2\right)
\eea
where
\bea\label{HKJ}
H &=& c_{\beta}^2 c_{\alpha}^2+g s_{\alpha}^2-fs_{\beta}^2 c_{\alpha}^2>0,\nn\\
K &=& f c_{\alpha}^2+g s_{\alpha}^2 (f c_{\beta}^2-s_{\beta}^2),\nn\\
J &=& H^{-1} \left[ g(c_{\beta}^2-f s_{\beta}^2)+K^{-1} (1-f)^2 c_{\beta}^2 s_{\beta}^2 s_{\alpha}^2 c_{\alpha}^4\right].
\eea
and we have used the shorthand notation
\be
c_{\alpha}=\cos\alpha,\qquad s_{\alpha}=\sin\alpha,\qquad c_{\beta}=\cosh\beta,\qquad s_{\beta}=\sinh\beta.
\ee
Performing dimensional reduction along the $z$ direction to type IIA theory yields
\bea
ds_{10}^2 &=& -H^{-7/8} K \left[ dt+K^{-1} (1-f) c_{\beta} s_{\beta} s_{\alpha} c_{\alpha}^2 d\psi\right]^2\nn\\ 
&+& H^{1/8} \left[ dx_1^2+\cdots +dx_5^2+J d\psi^2+A \left( \fft{dx^2}{G(x)}-\fft{dy^2}{G(y)}+B d\phi^2\right)
\right],\nn\\
F_\2 &=& c_{\beta} s_{\beta} c_{\alpha}\ d[H^{-1} (f-1)]\wedge  dt+c_{\alpha} s_{\alpha}\ d[H^{-1} (c_{\beta}^2-g-f s_{\beta}^2)]\wedge  d\psi,\nn\\
\phi &=& -\fft34 \log H.
\eea
This describes the superposition of a nonextremal D0-brane smeared along five directions, an F7-brane wrapped on a 3-sphere, and a KK bubble. This solution has angular momentum associated with the gravitational field that goes as $c\ c_{\beta} s_{\beta} s_{\alpha} c_{\alpha}^2$. The surface at $y=-\fft{1}{c}$ is the event horizon, just as it was in the seed solution. However, this solution contains an outer horizon as well, which is located at $K=0$. In analogy with the Kerr metric, the space between these two horizons is the ergoregion.

It is rather surprising that this solution has angular momentum, given the fact that the seed solution has none. However, being the outcome of the solution generation, the angular momentum is not an independent quantity, since it can be expressed in terms of the parameters $c$, $\alpha$ and $\beta$ which are related to the mass and charges. The angular momentum associated with the gravitational field, along with that associated with the electromagnetic field, keep the system in a stable configuration without the need of an applied external field. There may exist more general solutions for which the angular momentum is an independent quantity. However, such solutions are likely to be unstable, and would either slow down to the above configuration or else not have enough rotation to prevent their collapse.

Note that the angular momentum cannot be related to a brane charge via T-duality, since dualizing along the $\psi$ direction would lead to an ill-behaved solution with a singularity along $x=1$. However, we can T-dualize along the $x_1,\dots, x_5$ directions to get the type IIB solution
\bea\label{D5-Schwarzschild}
ds_{10}^2 &=& H^{-1/4}\left[ -K \left[ dt+K^{-1} (1-f) c_{\beta} s_{\beta} s_{\alpha} c_{\alpha}^2 d\psi\right]^2
+dx_1^2+\cdots +dx_5^2\right]\nn\\
&+& H^{3/4} \left[ J d\psi^2+A \left( \fft{dx^2}{G(x)}-\fft{dy^2}{G(y)}+B d\phi^2\right)\right],\nn\\
\ast F_\3 &=&  \left( c_{\beta} s_{\beta} c_{\alpha}\ d[H^{-1} (f-1)]\wedge  dt +c_{\alpha} s_{\alpha}\ d[H^{-1} (c_{\beta}^2-g-f s_{\beta}^2)]\wedge  d\psi\right) \wedge d^5x,\nn\\
\phi &=& -\fft12 \log H.
\eea
This describes a nonextremal D5-brane superimposed with a smeared F2-brane and a KK bubble, which has the same angular momentum as we had prior to performing the chain of T-dualities. 

In the limit $\alpha=0$, the F2-brane is removed and we are left with a nonextremal D5-brane and a KK bubble:
\be
ds_{10}^2 = H^{-1/4}\left( -f dt^2+dx_1^2+\cdots +dx_5^2\right)
+ H^{3/4} \left[ g d\psi^2+A \left( \fft{dx^2}{G(x)}-\fft{dy^2}{G(y)}+B d\phi^2\right)\right],
\ee
for which $F_\3$ and $\phi$ are the same as in (\ref{D5}) and
\be\label{H-0alpha}
H=c_{\beta}^2-f s_{\beta}^2.
\ee
On the other hand, taking the limit $\beta=0$ and T-dualizing back along the $x_1,\dots ,x_5$ directions yields
\bea
ds_{10}^2 &=& H^{-7/8} g d\psi^2+H^{1/8} \left[ -f dt^2+dx_1^2+\cdots +dx_5^2+A \left( \fft{dx^2}{G(x)}-\fft{dy^2}{G(y)}+B d\phi^2\right)\right] ,\nn\\
\eea
for which $F_\3$ and $\phi$ are the same as in (\ref{F7}) and 
\be\label{H-0beta}
H=c_{\alpha}^2+g s_{\alpha}^2.
\ee
This describes a neutral 5-brane superimposed with an F7-brane wrapped on a 2-sphere.

\section{Other examples}

\subsection{Superposition of two D5-branes and smeared F2-brane on KK bubble}

Two black holes on a KK bubble is described by the metric \cite{elvang}
\be\label{2bh}
ds_5^2=-f dt^2+g d\psi^2+(R_1-\zeta_1)(R_4+\zeta_4) d\phi^2+\fft{Y_{14} Y_{23}}{4R_1R_2R_3R_4} \sqrt{\fft{Y_{12} Y_{34}}{Y_{13} Y_{24}}} \left( \fft{R_1-\zeta_1}{R_4-\zeta_4}\right) (dr^2+dy^2),
\ee
where
\be
\zeta_i=y-c_i,\qquad R_i=\sqrt{r^2+\zeta_i^2},\qquad Y_{ij}=R_i R_j+\zeta_i \zeta_j+r^2,
\ee
and
\be
f=\fft{(R_2-\zeta_2)(R_4-\zeta_4)}{(R_1-\zeta_1)(R_3-\zeta_3)},\qquad g=\fft{R_3-\zeta_3}{R_2-\zeta_2}.
\ee
As opposed to a four-dimensional static solution describing two black holes held apart by a strut (conical singularity) \cite{israel,myers}, in this solution it is the KK bubble that serves to hold the black holes apart in static equilibrium. 
In fact, for $c_2<c_1,c_3$, a periodicity in $\phi$ of $2\pi$ and a periodicity in $\psi$ of
\be
\Delta\psi=\fft{8\pi c_2 (c_1+c_3)}{\sqrt{(c_1+c_2)(c_2+c_3)}},
\ee
the solution is regular and free of conical deficits. The two black hole horizons are located at $r=0$, $c_2<y<c_1$ and $r=0$, $-c_3<y<-c_2$. The geometry is asymptotically Minkowski$_4\times S^1$.

Embedding (\ref{2bh}) in eleven dimensions with (\ref{general11Dmetric}), performing the boost (\ref{boost}) and the rotation (\ref{rotation}), dimensionally reducing along the boosted and rotated $z$ direction and then T-dualizing along the $x_1,\dots ,x_5$ directions yields the superposition of two nonextremal D5-branes and a smeared F2-brane on a KK bubble:
\bea
ds_{10}^2 &=& H^{-1/4}\left[ -K \left[ dt+K^{-1} (1-f) c_{\beta} s_{\beta} s_{\alpha} c_{\alpha}^2 d\psi\right]^2
+dx_1^2+\cdots +dx_5^2\right]\\
&+& H^{3/4} \left[ J d\psi^2+(R_1-\zeta_1)(R_4+\zeta_4) d\phi^2+\fft{Y_{14} Y_{23}}{4R_1R_2R_3R_4} \sqrt{\fft{Y_{12} Y_{34}}{Y_{13} Y_{24}}} \left( \fft{R_1-\zeta_1}{R_4-\zeta_4}\right) (dr^2+dy^2)\right],\nn
\eea
where $F_\3$ and $\phi$ are given by (\ref{D5-Schwarzschild}) and $H$, $K$ and $J$ are given by (\ref{HKJ}). This solution has angular momentum that is fixed in terms of the mass and magnetic charges.

\subsection{Superposition of D5-brane and smeared F2-brane on Euclidean Kerr instanton}

A static black hole on the Euclidean Kerr instanton was constructed in \cite{ford}, and can be expressed in C-metric-like coordinates \cite{teo1} as
\bea\label{Kerr}
ds_5^2 &=& -f dt^2+g (d\psi+\Omega)^2
+ A\left( \fft{dx^2}{G(x)}-\fft{dy^2}{G(y)}+B d\phi^2\right),
\eea
where $G(x)$ and $f$ are given by (\ref{Gf}), $g=C/D$ and
\bea
\Omega &=& \fft{2\alpha c^2 \chi^2 [1+c-(1-c)\alpha^2][1-c-(1+c)\alpha^2]}{(1+\alpha^2)}\nn\\
&\times& \fft{(1+y)[(1-y)(2-c+cx)+(1-x)(2-c+cy)\alpha^2]G(x)}{(1-x)(x-y)F(x,y)}\ d\phi,\nn\\
A &=& \fft{2\chi^4 (1+cx)K(x,y)}{c^2 (1-c)(1-\alpha^2)(1+\alpha^2)^2 (x-y)^3},\nn\\
B &=& -\fft{2c^2(1-c)(1-\alpha^2)(1+\alpha^2)^2 G(x) G(y)}{(x-y)(1+cy)F(x,y)},\nn\\
C &=& \fft{c^2 (1-\alpha^2)(1+cx)}{1+c} \big[(1-c)(1-x)(1-y)\big(1-c-(1+c)\alpha^2\big)\big(1+c-(1-c)\alpha^2\big)\nn\\
&-& 8\alpha^2(c+x+y+cxy)\big],\nn\\
D &=& (1+cx)\left[ (1-c)\left(1+c-(1-c)\alpha^2\right)-(1+cy)\left(1-c-(1+c)\alpha^2\right) \right] ^2\nn\\
&-& \alpha^2 (1+cy) \left[ (1-c)\left( 1-c-(1+c)\alpha^2\right)-(1+cx)\left( 1+c-(1-c)\alpha^2\right)\right]^2.
\eea
The parameters $\chi$, $c$ and $\alpha$ take the ranges $\chi>0$, $0\le c<1$ and $\alpha^2<\fft{1-c}{1+c}$, and the $x$ and $y$ coordinates take the ranges $-1\le x\le 1$ and $-\fft{1}{c}\le y\le -1$. The horizon is at $y=-\fft{1}{c}$ and the asymptotic region is at $x=y=-1$. There are no singularities or closed timelike curves outside of the horizon. The horizon topology is $S^3$ and the asymptotic geometry is Minkowski$_4\times S^1$. While the $S^1$ generally blows up at infinity, it remains finite if the following combination of parameters
\be
\fft{\sqrt{1-c^2}(1+\alpha^2)}{\alpha},
\ee
is rational \cite{hunter}. The black hole on the Euclidean Schwarzschild instanton \cite{emparan} is recovered for vanishing $\alpha$.

Embedding (\ref{Kerr}) in eleven dimensions with (\ref{general11Dmetric}), performing the boost (\ref{boost}) and the rotation (\ref{rotation}), dimensionally reducing along the boosted and rotated $z$ direction and then T-dualizing along the $x_1,\dots ,x_5$ directions yields a nonextremal D5-brane on a Euclidean Kerr instanton superimposed with a smeared F2-brane:
\bea\label{solution-form1}
ds_{10}^2 &=& H^{-1/4}\left[ -K \left[ dt+K^{-1} (1-f) c_{\beta} s_{\beta} s_{\alpha} c_{\alpha} \left( c_{\alpha} d\psi+g \Omega\right)\right]^2
+dx_1^2+\cdots +dx_5^2\right]\nn\\
&+& H^{3/4} \left[ J (d\psi+L\Omega)^2+M\Omega^2+A \left( \fft{dx^2}{G(x)}-\fft{dy^2}{G(y)}+B d\phi^2\right)\right],\nn\\
\ast F_\3 &=&  \left( c_{\beta} s_{\beta} c_{\alpha}\ d[H^{-1} (f-1)]\wedge  dt +c_{\alpha} s_{\alpha}\ d[H^{-1} (c_{\beta}^2-g-f s_{\beta}^2)]\wedge  d\psi\wedge -s_{\alpha} d[H^{-1} g\Omega]\right) \wedge d^5x,\nn\\
\phi &=& -\fft12 \log H.
\eea
where $H$, $K$ and $J$ are given by (\ref{HKJ}) and
\bea\label{LM}
L &=& \left[ g-(c_{\beta}^2-g-f s_{\beta}^2) s_{\alpha}+H^{-1} K^{-1} (1-f)^2 g c_{\beta}^2 s_{\beta}^2 c_{\alpha}^2 s_{\alpha}^2\right] J^{-1} c_{\alpha},\nn\\
M &=& K^{-1} fg c_{\alpha}^2-J L^2.
\eea
This solution has angular momentum that is fixed in terms of the mass and magnetic charges.

\subsection{Superposition of D5-brane and smeared F2-brane on Taub-bolt instanton}

A static black hole on the non-self-dual Taub-NUT instanton was obtained in \cite{ford} and can be expressed in C-metric-like coordinates \cite{teo1} by the metric (\ref{Kerr})
where $G(x)$ and $f$ are given by (\ref{Gf}), $g=C/D$ and
\bea
\Omega &=& \fft{2\alpha\chi^2 [2+x+y+c(1+x)(1+y)]}{(1-\alpha^2)(x-y)}d\phi,\nn\\
A &=& \fft{2\chi^4 (1-c)(1+cx) K(x,y)}{(1-\alpha^2)(x-y)^3},\nn\\
B &=& -\fft{2(1+x)(1+y)}{(1-c)(x-y)},\nn\\
C &=& (1-\alpha^2)(1-x)(1-y)(1+cx),\nn\\
D &=& (1+cx)(1-y)^2-\alpha^2 (1+cy)(1-x)^2.
\eea
The parameters $\chi$, $c$ and $\alpha$ take the ranges $\chi>0$, $0\le c<1$ and $\alpha^2\ge 1$, and the $x$ and $y$ coordinates take the ranges $-1\le x\le 1$ and $-\fft{1}{c}\le y\le -1$. The horizon is located at $y=-\fft{1}{c}$ 
with a topology of $S^3$. The asymptotic region is at $x=y=-1$ and is a nontrivial finite $S^1$ fiber bundle over Minkowski$_4$. The solution has no singularities or closed timelike curves outside of the horizon for either $\alpha^2=(1-c^2)/4$ or $\alpha=1$, the latter being the case of a black hole on the self-dual Taub-NUT instanton \cite{ishihara}. On the other hand, for vanishing $\alpha$ the black hole on the Euclidean Schwarzschild instanton \cite{emparan} is recovered.

Embedding (\ref{Kerr}) in eleven dimensions with (\ref{general11Dmetric}), performing the boost (\ref{boost}) and the rotation (\ref{rotation}), dimensionally reducing along the boosted and rotated $z$ direction and then T-dualizing along the $x_1,\dots ,x_5$ directions yields a nonextremal D5-brane on a Taub-bolt instanton superimposed with a smeared F2-brane. The solution has the form (\ref{solution-form1}) where $H$, $K$ and $J$ are given by (\ref{HKJ}) and $L$ and $M$ are given by (\ref{LM}). This solution has angular momentum that is fixed in terms of the mass and magnetic charges.

\subsection{D5-brane on Eguchi-Hanson instanton}

The metric for a rotating black hole on the Eguchi-Hanson instanton is contained in the rotating black lens solution found it \cite{teo2}. In C-metric-like coordinates, the metric for the case in which the black hole has a single angular momentum can be written as
\bea\label{Eguchi-Hanson}
ds_5^2 &=& -\fft{H(y,x)}{H(x,y)} \left( dt-\omega_{\psi}d\psi-\omega_{\phi}d\phi\right)^2-\fft{F(x,y)}{H(y,x)}d\psi^2+\fft{2J(x,y)}{H(y,x)} d\psi d\phi\nn\\
&+& \fft{F(y,x)}{H(y,x)} d\phi^2+\fft{\chi^2 H(x,y)}{2(1-a^2)(1-b)^3 (x-y)^2} \left( \fft{dx^2}{G(x)}-\fft{dy^2}{G(y)}\right),
\eea
where
\bea
\omega_{\psi} &=& \fft{2\chi}{H(y,x)} \sqrt{\fft{2b(1+b)(b-c)}{(1-a^2)(1-b)}} (1-c)(1+y) \{2[1-b-a^2(1+bx)]^2(1-c)\nn\\
&-& a^2(1-a^2)b(1-b)(1-x)(1+cx)(1+y)\},\nn\\
\omega_{\phi} &=& \fft{2\chi}{H(y,x)} \sqrt{\fft{2b(1+b)(b-c)}{(1-a^2)(1-b)}} a(1-c)(1+x)^2 (1+y)[a^4(1+b)(b-c)\nn\\
&+& a^2 (1-b)(b(c-1)+2c)-(1-b)^2c],\nn\\
G(x) &=& (1+cx)(1-x^2),\nn\\
H(x,y) &=& 4(1-b)(1-c)(1+bx)\{ (1-b)(1-c)-a^2[(1+bx)(1+cy)+(b-c)(1+y)]\}\nn\\
&+& a^2(b-c)(1+x)(1+y)\{(1+b)(1+y)[(1-a^2)(1-b)c(1+x)+2a^2b(1-c)]\nn\\
&-& 2b(1-b)(1-c)(1-x)\},\nn\\
F(x,y) &=& \fft{2\chi^2}{(1-a^2)(x-y)^2} \Big[ 4(1-c)^2(1+bx)[1-b-a^2(1+bx)]^2 G(y)\nn\\
&-& a^2 G(x) (1+y)^2 \Big( [1-b-a^2(1+b)]^2(1-c)^2(1+by)-(1-a^2)(1-b^2)\nn\\
&\times& (1+cy)\{ (1-a^2)(b-c)(1+y)+[1-3b-a^2(1+b)](1-c)\}\Big)\Big],
\eea
\bea
J(x,y) &=& \fft{4\chi^2 a(1-c)(1+x)(1+y)}{(1-a^2)(x-y)} [1-b-a^2(1+b)][(1-b)c+a^2(b-c)]\nn\\
&\times& [(1+bx)(1+cy)+(1+cx)(1+by)+(b-c)(1-xy)].\nn
\eea
The parameters $\chi$ and $c$ take the ranges $\chi>0$ and $0\le c<1$ and the parameters $a$ and $b$ are fixed  as
\be
a=\fft{3(1-c)}{3+5c},\qquad b=\fft{4c(3-c)}{5c^2-6c+9}.
\ee
The $x$ and $y$ coordinates take the ranges $-1\le x\le 1$ and $-\fft{1}{c}\le y\le -1$. The horizon is at $y=-\fft{1}{c}$ and has topology $S^3$. The region outside the horizon does not contain any singularities and no closed timelike curves have been found \cite{teo1}. The asymptotic region at $x=y=-1$ is Minkowski$_5/{\mathbb Z}_2$. For vanishing $c$, the direct product of the Eguchi-Hanson instanton with time is recovered.

Since there is no $S^1$ that remains finite in the asymptotic region, we cannot construct a well-behaved solution involving the $\alpha$ parameter associated with the rotation (\ref{rotation}), though we can still consider one with the $\beta$ parameter associated with the boost (\ref{boost}). Embedding (\ref{Eguchi-Hanson}) in eleven dimensions with (\ref{11D-metric1}), performing the boost (\ref{boost}), dimensionally reducing along the boosted $z$ direction and T-dualizing along the $x_1,\dots ,x_5$ directions yields a nonextremal D5-brane on the Eguchi-Hanson instanton:
\bea
ds_{10}^2 &=& H^{-1/4}\left( -\fft{K(y,x)}{K(x,y)} (dt-\omega_{\psi} d\psi-\omega_{\phi}d\phi)^2 +dx_1^2+\cdots +dx_5^2\right)
+H^{3/4} ds_4^2,\nn\\
\ast F_\3 &=& \coth\beta\ dH^{-1}\wedge dt\wedge d^5x+\sinh\beta \left[ d\left( \fft{K(y,x)\omega_{\phi}}{HK(x,y)}\right) \wedge d\phi+d\left( \fft{K(y,x)\omega_{\psi}}{HK(x,y)}\right)\wedge d\psi\right]\wedge d^5x,\nn\\
\phi &=& -\fft12 \log H,
\eea
where
\be
H=1+\left( 1-\fft{K(y,x)}{K(x,y)}\right) \sinh^2\beta.
\ee
%

\section*{Acknowledgments}

This work is supported in part by NSF grant PHY-0969482 and a PSC-CUNY Award jointly funded by The Professional Staff Congress and The City University of New York.

\end{document}